\documentclass[trackchanges, twocolumn]{aastex701}
\usepackage{graphicx}
\usepackage{txfonts}
\usepackage{xcolor}
\usepackage{lastpage}
\usepackage{xspace}

\newcommand{\BGP}{\texttt{BGP}\xspace}
\newcommand{\Teff}{\ifmmode {T_{\rm eff}}\else${T_{\rm eff}}$\fi}
\newcommand{\Msun}{\ensuremath{\mathrm{M}_\odot}\xspace}

\newcommand{\Rsun}{\ensuremath{R_\odot}}

\newcommand{\Zsun}{\ensuremath{Z_\odot}\xspace}
\newcommand{\Mdonor}{\ensuremath{M_\mathrm{donor}}\xspace}
\newcommand{\MBH}{\ensuremath{M_\mathrm{BH}}\xspace}
\newcommand{\chieff}{\ensuremath{\chi_\mathrm{eff}}\xspace}

\newcommand{\posydon}{\texttt{POSYDON}\xspace}
\newcommand{\mesa}{\texttt{MESA}\xspace}

\newcommand{\Mone}{\ensuremath{\mathrm{M}_{1}}\xspace}

\begin{document}

\title{High-mass binary black hole mergers from detailed binary evolution models}

\author[orcid=0000-0002-6842-3021,sname='Briel']{Max M. Briel}
\affiliation{Département d'Astronomie, Université de Genève, Chemin Pegasi 51, CH-1290 Versoix, Switzerland}
\affiliation{Gravitational Wave Science Center (GWSC), Université de Genève, CH-1211 Geneva, Switzerland}
\email[show]{max.briel@gmail.com}

\author[orcid=0009-0004-6457-9530,sname='Bıyıklı']{Olcay Bıyıklı}
\affiliation{Department of Physics, Bilkent University, Bilkent, 06800, Ankara, Turkey}
\email[]{}

\author[orcid=0000-0003-1474-1523,sname='Fragos']{Tassos Fragos}
\affiliation{Département d'Astronomie, Université de Genève, Chemin Pegasi 51, CH-1290 Versoix, Switzerland}
\affiliation{Gravitational Wave Science Center (GWSC), Université de Genève, CH-1211 Geneva, Switzerland}
\email[]{}

\author[orcid=0000-0002-7322-4748,sname='Ray']{Anarya Ray}
\affiliation{Center for Interdisciplinary Exploration and Research in Astrophysics (CIERA), Northwestern University, 1800 Sherman Ave, Evanston, IL 60201, USA}
\affiliation{The NSF-Simons AI Institute for the Sky (NSF-Simons SkAI), 172 E. Chestnut Street, Chicago, IL 60611, USA}
\email[]{}

\author[orcid=0000-0002-0031-3029,sname='Xing']{Zepei Xing}
\affiliation{Center for Interdisciplinary Exploration and Research in Astrophysics (CIERA), Northwestern University, 1800 Sherman Ave, Evanston, IL 60201, USA}
\email[]{}

\author[orcid=0000-0003-0648-2402,sname='Gallegos-Garcia']{Monica Gallegos-Garcia}
\affiliation{Center for Interdisciplinary Exploration and Research in Astrophysics (CIERA), Northwestern University, 1800 Sherman Ave, Evanston, IL 60201, USA}
\affiliation{Department of Physics and Astronomy, Northwestern University, 2145 Sheridan Road, Evanston, IL 60208, USA}
\affiliation{Center for Astrophysics \textbar{} Harvard \& Smithsonian, 60 Garden St. Cambridge, MA, 02138, USA}
\affiliation{Harvard Society of Fellows, 78 Mount Auburn Street, Cambridge, MA 02138, USA}
\email[]{}

\author[orcid=0000-0002-6064-388X,sname='Chattaraj']{Abhishek Chattaraj}
\affiliation{Department of Physics, University of Florida, 2001 Museum Rd, Gainesville, FL 32611, USA}
\email[]{}

\author[orcid=0000-0001-5261-3923,sname='Andrews']{Jeff J. Andrews}
\affiliation{Department of Physics, University of Florida, 2001 Museum Rd, Gainesville, FL 32611, USA}
\affiliation{Institute for Fundamental Theory, 2001 Museum Rd, Gainesville, FL 32611, USA}
\email[]{}

\author[orcid=0000-0002-0147-0835,sname='Zevin']{Michael Zevin}
\affiliation{The Adler Planetarium, 1300 South DuSable Lake Shore Drive, Chicago, 60605, IL, USA}
\affiliation{Center for Interdisciplinary Exploration and Research in Astrophysics (CIERA), Northwestern University, 1800 Sherman Ave, Evanston, IL 60201, USA}
\affiliation{The NSF-Simons AI Institute for the Sky (NSF-Simons SkAI), 172 E. Chestnut Street, Chicago, IL 60611, USA}
\email[]{}

\author[orcid=0000-0001-9236-5469,sname='Kalogera']{Vicky Kalogera}
\affiliation{Center for Interdisciplinary Exploration and Research in Astrophysics (CIERA), Northwestern University, 1800 Sherman Ave, Evanston, IL 60201, USA}
\affiliation{The NSF-Simons AI Institute for the Sky (NSF-Simons SkAI), 172 E. Chestnut Street, Chicago, IL 60611, USA}
\affiliation{Department of Physics and Astronomy, Northwestern University, 2145 Sheridan Road, Evanston, IL 60208, USA}
\email[]{}

\author[orcid=0000-0001-6692-6410,sname='Gossage']{Seth Gossage}
\affiliation{Center for Interdisciplinary Exploration and Research in Astrophysics (CIERA), Northwestern University, 1800 Sherman Ave, Evanston, IL 60201, USA}
\affiliation{The NSF-Simons AI Institute for the Sky (NSF-Simons SkAI), 172 E. Chestnut Street, Chicago, IL 60611, USA}
\email[]{}

\author[orcid=0000-0003-1749-6295,sname='Srivastava']{Philipp M. Srivastava}
\affiliation{Center for Interdisciplinary Exploration and Research in Astrophysics (CIERA), Northwestern University, 1800 Sherman Ave, Evanston, IL 60201, USA}
\affiliation{The NSF-Simons AI Institute for the Sky (NSF-Simons SkAI), 172 E. Chestnut Street, Chicago, IL 60611, USA}
\affiliation{Electrical and Computer Engineering, Northwestern University, 2145 Sheridan Road, Evanston, IL 60208, USA}
\email[]{}

\author[orcid=0000-0003-0420-2067,sname='Teng']{Elizabeth Teng}
\affiliation{Center for Interdisciplinary Exploration and Research in Astrophysics (CIERA), Northwestern University, 1800 Sherman Ave, Evanston, IL 60201, USA}
\affiliation{The NSF-Simons AI Institute for the Sky (NSF-Simons SkAI), 172 E. Chestnut Street, Chicago, IL 60611, USA}
\affiliation{Department of Physics and Astronomy, Northwestern University, 2145 Sheridan Road, Evanston, IL 60208, USA}
\email[]{}
 
\begin{abstract}
Gravitational-wave observations reveal a population of binary black hole (BBH) mergers with primary masses above ${\sim}40\,\Msun$, extending into and potentially beyond the pair-instability mass gap, with a possibly flat mass-ratio and broader \chieff distribution.
We investigate whether super-Eddington accretion during stable mass transfer in isolated binary evolution can produce BBH mergers consistent with these properties across primary BH mass, mass-ratio, and \chieff distributions.
Using \posydon, we simulate BBH merger populations with primary BH masses above ${\sim}40\Msun$, under three BH accretion efficiencies: Eddington-limited, GRRMHD-informed, and fully conservative. We additionally vary the natal kick strength, including strong kicks at high BH masses.

We find that super-Eddington accretion does not suppress BBH mergers in the high-mass regime. Fully-conservative accretion leads to an increase of BBH mergers in \posydon with a strong kick-independent peak at $\chieff=0.6$ and a sharp mass-ratio peak at $q\sim0.5$, whereas observations favor $\chieff=0.0$ and a flatter mass-ratio distribution. The GRRMHD-informed and Eddington-limited accretion are compatible with the observed primary BH mass and mass ratio distribution, but require natal kicks to populate negative \chieff.

A joint analysis of the primary BH mass, mass ratio, and \chieff distributions provides strong constraints on binary evolution physics, and disfavor fully-conservative BH accretion as the dominant formation mechanism for high-mass BBH mergers. The Eddington-limited and GRRMHD-informed prescriptions with modest kicks can explain part of the high-mass population, but an additional formation channel is still needed to account for the high fraction of negative \chieff systems and high secondary BH spins.

\end{abstract}

%

\section{Introduction} \label{sec:introduction}

Very massive stars ($\mathrm{M}_\mathrm{ZAMS}\approx100{-}200\,\Msun$) are expected to undergo electron-positron pair production during late evolutionary stages, triggering a collapse followed by explosive oxygen burning \citep{Fowler+64, Barkat+67}. 
Depending on the stellar mass, this process can drive violent pulsations that strip the hydrogen envelope and part of the helium core, known as pulsational pair-instability \citep{Woosley+17} or result in the complete disruption of the star in a pair-instability supernova \citep[PISN;][]{Rakavy+67, Fraley+68, Woosley+02}.
Consequently, this mechanism is thought to prevent the direct formation of black holes (BHs) with masses in the range of ${\sim}50{-}130\,\Msun$, giving rise to the so-called PISN or upper BH mass gap. The precise boundaries of this gap are sensitive to the adopted stellar physics, such as nuclear reaction rates \citep{Farmer+19, Farag+22}, mass loss \citep{Winch+24}, and rotation \citep{Winch+25} with the lower mass shifting between ${\sim}40\,\Msun$ and ${\sim}70\,\Msun$ depending on these assumptions. At stellar masses above the PISN regime, BHs are able to form through direct collapse driven by photo-disintegration \citep{Heger+03}.

Gravitational-wave observations from the LIGO--Virgo--KAGRA (LVK) detector network \citep{Acernese+14, Abbott+20d, Akutsu+21} have revealed a growing population of binary black hole (BBH) mergers with BH masses above $40\,\Msun$, with over 100 of such events identified through \texttt{GWTC-5.0} \citep{Abac+25a, Abac+25b, Abac+26}. Among these, GW190521 \citep{Abbott+20b} and GW231123 \citep{Abac+25c} stand out as the most massive BBH mergers detected to date, with primary BH masses of $85^{+21}_{-14}\,\Msun$ and $137^{+23}_{-18}\,\Msun$, respectively, placing both systems within the expected PISN mass gap. GW231123, additionally, contains a secondary BH with a mass of $101^{+22}_{-50}\,\Msun$.

Recent population inference further suggests that these high-mass BBH mergers have a flatter mass ratios distribution \citep{Banagiri+25, Ray+25a} or a preference for unequal mass ratios \citep{Ray+26}. This population also exhibit a broader \chieff distribution compared to the lower-mass population \citep{Abac+26a}, with evidence for a subpopulation of BHs with high individual spins \citep[up to $\chi\sim0.7$;][]{Antonini+25, Banagiri+25}. This transition in properties occurs around $\sim45\Msun$, which is near the expected lower boundary of the PISN mass gap. While some studies have interpreted the location and properties of this transition as evidence that the high-mass population originates from hierarchical mergers \citep{Antonini+25}, other studies show that the current spin and mass distributions remain too uncertain to reach this conclusion without relying on strong model assumptions \citep{Wang+25a, Ray+25a, Wolfe+26, Abac+26a}. Still other studies find evidence of spin alignment in the high-mass regime \citep{Rinaldi+26, Li+26}. For example, under a separate set of model assumptions, \citet{Wang+25a} identify an additional population of low-spin massive BHs ($\Mone\gtrsim60\Msun$), which shifts the transition mass upward to $68.5^{+19.8}_{-18.5}\Msun$. Although consistent within error bars, this illustrates the sensitivity to model choices. Although the exact formation mechanism remains under debate, current observations consistently point to a change in spin and mass ratio distribution near the expected lower edge of the PISN mass gap.

While the primary BH mass distribution does not exhibit a clear PISN mass gap, \citet{Tong+25} report a sharp cut-off in the less massive (secondary) BH mass distribution in \texttt{GWTC-4.0}. Using a more flexible model on the same catalog, however, \citet{Ray+25a} find no statistical evidence for a sharp cut-off. This is further confirmed by \citet{Abac+26a} with \texttt{GWTC-5.0}, disfavoring a sharp mass gap but finding a suppression of high $\mathrm{M}_2$. This suggests that the secondary BHs more closely reflect their stellar progenitor masses, preserving the PISN mass gap, while the primary BHs may have been shifted into the gap through processes, such as super-Eddington mass transfer \citep[e.g.,][]{vanSon+20, Briel+23} or hierarchical mergers \citep[e.g.,][]{Schnittman+07, Baker+07}. The contrasting behavior of primary and secondary BHs raises the question of which formation channels can produce high-mass, potentially high-spin primaries while leaving secondaries below the PISN mass gap. This behavior excludes alterations in the nuclear reaction rates, because variations to the rates will affect both the primary and secondary mass distributions equally by shifting the PISN mass range \citep{Farmer+19, Farag+22}. Instead, a mechanism that preferentially boosts the primary BH mass is required. 

Hierarchical mergers in dense star clusters can in principle produce this asymmetry, as mergers between a lower-mass first-generation and a higher-mass second-generation BH can yield a primary mass in the PISN gap. However, the isotropic distribution of individual BH spin vectors in cluster environments makes it challenging to produce the very high spins ($\chi>0.8$) inferred for PISN mass gap events, such as GW231123 \citep{Abac+25c} through dynamical mergers alone \citep{Rodriguez+19, Stegmann+25}. Nonetheless, the generally flatter spin and mass ratio distributions of mergers above $\Mone>40\Msun$ could indicate a dynamical origin. Besides hierarchical mergers, additional processes for the primary to gain mass, such as coherent misaligned gas accretion \citep{Bartos+25} or primordial BHs \citep[e.g.][]{Carr+75, Escriva+23, Afroz+25b, DeLuca+26}, have been proposed as pathways to populate the PISN mass gap. Independent of the exact origin of this asymmetry, population inference in the $40{-}60\Msun$ regime indicates comparable contributions from isolated binaries and hierarchical mergers \citep{Ray+26}.

Within the context of isolated binary evolution, primary BHs can reach masses within the PISN mass gap through two distinct pathways without requiring a shift in the PISN boundaries. First, at sub-solar metallicities, even massive stars that undergo mass transfer can retain substantial hydrogen envelopes until core-collapse \citep{Gotberg+17, Klencki+20}, as long as pair-instability is avoided \citep{Woosley+17}. During direct collapse, this envelope falls back onto the newly formed BH rather than being ejected, producing a more massive BH, potentially in the PISN mass gap. In single stars, the inclusion of the hydrogen envelope has been shown to produce BHs up to ${\sim}93\Msun$, though the exact upper limit depends on the wind mass loss prescription, core overshooting, rotation and lower PISN limit \citep{Woosley+17, Winch+24}.

Secondly, super-Eddington accretion onto the first-born BH during stable mass transfer (SMT) can efficiently increase its mass and spin into the PISN mass gap \citep{Shao+22}. Since SMT is expected to form more massive BHs \citep{Neijssel+19, vanSon+22a} than other isolated binary evolution sub-channels, such as common envelope evolution, this represents an efficient channel for boosting the primary BH mass. Earlier work suggested that conservative accretion suppresses BBH systems from merging within the Hubble time, as reduced angular momentum loss leads to less orbital shrinkage compared to Eddington-limited accretion \citep{vanSon+20, Bavera+21}. However, detailed binary models with fully-conservative accretion onto the BH still lead to the formation of BBH mergers with the primary mass inside the PISN mass gap without a significant reduction in the overall BBH merger rate \citep{Briel+23}. This discrepancy lies in the treatment of Case A mass transfer, which is the dominant interaction in the SMT channel for BBH mergers in \posydon \citep{Briel+26}. We discuss this in more detail in Section \ref{sec:accretion} and Appendix \ref{app:super_eddington}. Similarly, for the Eddington-limited case, \citet{Klencki+26} show that with detailed treatment of the donor star during the BH accretion phase, the minimum final orbital configuration post mass transfer does not change for different angular momentum loss prescription. These results suggest that super-Eddington accretion does not suppress the overall BBH merger rate from the SMT channel and can efficiently boosts the primary BH mass to populate the PISN mass gap.

In this work, we perform a population synthesis study of the local BBH merger population with primary BH masses above $40\Msun$ using \posydon \citep{Fragos+23, Andrews+25}, allowing us to test the effect of super-Eddington accretion using large pre-computed grids of detailed binary models run with \mesa \citep{Paxton+11, Paxton+13, Paxton+15, Paxton+18, Paxton+19, Jermyn+23}. We include full fallback of the hydrogen envelope at core-collapse and explore three BH accretion efficiencies to investigate whether isolated binary evolution can populate the PISN mass gap in the primary BH mass distribution and reproduce the multi-dimensional properties of the observed high-mass BBH population. We further examine the role of natal kick strength in shaping these distributions and compare them against the inferred intrinsic population of BBH mergers with $\Mone>39.7\,\Msun$ from the data-driven binned gaussian process (\BGP) model from \texttt{GWTC-5.0} \citep{Abac+26a}.
We present the population synthesis setup in Section \ref{sec:methods}.
In Section \ref{sec:results}, we present and discuss these populations in terms of primary BH mass, mass ratio and effective spin parameter. Individual spin components and predicted tilts are examined in Section \ref{sec:discussion}, where we also place our finding in the context of the broader literature. We summarize our conclusions in Section \ref{sec:conclusion}.

\section{Method} \label{sec:methods}

We simulate our synthetic BBH merger populations using \posydon v2.1.6\footnote{\posydon releases are publicly available on \href{https://github.com/POSYDON-code/POSYDON/releases}{Github}.} \citep{Fragos+23, Andrews+25}. \posydon uses pre-computed grids of detailed binary-evolution models, run with the \mesa stellar-structure and binary-evolution code, to evolve binaries from the zero-age main-sequence (ZAMS) to a gravitational wave merger across eight metallicities between $10^{-4}\,\Zsun$ and $2\,\Zsun$, assuming $\Zsun=0.0142$. 

For different evolutionary phases, \posydon contains unique binary-evolution model grids. For binaries with two stellar components, it uses the \texttt{HMS-HMS} grid. When the donor star is hydrogen-rich and the companion is a compact object (BH or neutron star), \posydon uses the \texttt{CO-HMS\_RLO} grid, which starts at Roche lobe overflow \citep[for details, see][]{Fragos+23}. \posydon also includes \texttt{CO-HeMS} grids for helium-rich stars with a compact object companion, but most BBH progenitors discussed in this work do not evolve through these grids. By default, accretion onto the BH is limited to the Eddington-limited rate in \posydon. To determine the impact of the accretion efficiency onto BHs, additional detailed binary-star models have been computed,
one which allows for super-Eddington accretion informed by general relativistic radiative magnetohydrodynamic (GRRMHD-informed) simulations and a second fully-conservative accretion, following the simulation setup described by \citet{Xing+25}, but expanded to all eight metallicities modeled here. In all three accretion prescriptions, the angular momentum accreted onto the BH is computed at the innermost stable circular orbit (ISCO) following \citet{Thorne+74}.
In the GRRMHD-informed grids, the accretion efficiency is based on a series of GRRMHD simulations of magnetically arrested disks from Kwan et al. (in-prep), from which a fit for the relation between the accretion efficiency and the mass transfer is derived \citep[See, equation 1 in ][]{Xing+25}. Generally, this results in a BH accretion efficiency between 10\% to 30\%.

\subsection{Remnant mass prescription}
We adopt the \citet{Fryer+12}-delayed supernova remnant mass prescription, assuming the full fallback of the hydrogen envelope in the remnant mass calculation, but we find that other prescriptions do not appreciably change the high-mass BBH merger regime considered in this work. The inclusion of the hydrogen envelope increases both the mass and angular momentum deposited into the remnant, representing an optimistic upper bound on the maximum BH mass from direct collapse. The natal BH spin is estimated by following the collapse of the stellar profile at core carbon depletion onto a proto-BH of $2.5\,\Msun$, accounting for the angular momentum of the infalling material \citep[appendix D in][; see also section 8.3.4 in \citealt{Fragos+23}]{Bavera+21}. The effect of excluding the hydrogen envelope from the remnant mass calculation is discussed in Appendix \ref{app:no_H}.

For the pair-instability regime, we use an adapted version of the prescription described by \citet{Hendriks+23}, tuned to the maximum BH mass and PISN boundary from \citet{Farag+22}. Crucially, \citet{Farag+22}, in their work, do not show a shift in the PISN limits, but identifies a shift toward higher helium core masses at which pulsational pair-instability (PPI) set in, thanks to an improved simulation resolution compared to earlier work. We implement this by adopting $\Delta M_\mathrm{PPI}=-20\Msun$ and $M_\mathrm{CO}=0$ in the \citet{Hendriks+23} prescription, which places the onset of PPI at $M_\mathrm{He}\approx60\Msun$ and the lower PISN limit at $M_\mathrm{He}\approx 70\Msun$ \citep[See section 2.2.1 and figure 1 in][]{Andrews+25}.
For stars entering the PISN regime, we assume that no remnant is formed. For stars entering the PPI regime, the hydrogen envelope is removed and the pre-collapse mass loss is calculated based on the helium core size following equation 6 in \citet{Hendriks+23}. The pulsations in the PPI regime are generally sufficiently strong to eject the hydrogen envelope, which has a low binding energy when PPI conditions are met \citep{Woosley+17}. As a result of the higher helium core mass threshold for PPI onset, the lower edge of the PISN mass gap shifts to $\MBH\approx60\Msun$ and the build-up before the gap is less pronounced.

\subsection{Natal kick prescriptions}
During core-collapse, asymmetries in the ejected material can impart a natal momentum kick onto the newly formed compact object \citep[e.g.][]{Janka+94, Burrows+95, vandenHeuvel+97, Janka+24, Burrows+25}. In the regime of full fallback ($M_\mathrm{CO}\gtrsim11\Msun$ for \citet{Fryer+12}-delayed prescription), no material is ejected and only a recoil of a few km/s from the neutrino mass loss is expected \citep{Blaauw+61, Janka+24}. BHs with $M_1>39.7\,\Msun$ are, therefore, expected to receive negligible natal kicks. As such, we start our analysis with populations assuming no natal kick.

Nevertheless, some massive BBH mergers have been observed with misaligned spins compared to their orbit \citep[e.g.][]{Abac+25c, Alvarez-Lopez+26, Abac+26a}. Within isolated binary evolution, such misalignment will most likely arise from a natal kick, provided that mass loss during the life of its progenitor is isotropic, although spin tumbling during the supernova provides an additional mechanism for off-axis spin vectors \citep{Tauris+22}.
Precise astrometric observations of Galactic BH X-ray binaries \citep[e.g.][and references therein]{Miller-Jones+14} have enabled studies that robustly constrain the possible natal kicks imparted onto these BHs, both on an individual, system-per-system basis \citep[e.g.,][]{Willems+05,Gualandris+05,Fragos+09,Wong+12,Wong+14,Andrews+22, Kimball+23,DashwoodBrown+24} and at a population level \citep[e.g.,][]{Fragos+10,Repetto+12,Mandel+16a,Atri+19}. These studies have shown that X-ray binaries containing lower-mass BHs ($\lesssim 10\,\rm M_{\odot}$) require natal kicks in excess of $\gtrsim 80\,\rm km\,s^{-1}$ to explain their current properties. On the other hand, for higher-mass BHs ($\gtrsim 10\,\rm M_{\odot}$) found in high-mass X-ray binaries, which are thought to be more closely connected to the evolution of GW source progenitors, one can only set an upper limit on the possible natal kick that has been imparted on the BH during its formation, indicating natal kicks $\lesssim 100\,\rm km\,s^{-1}$. For the latter class of systems, more stringent conclusions are hampered by the small number of systems for which this type of analysis is possible. For BHs in the mass range of $\gtrsim 40\,\rm M_{\odot}$ no constraints are available. Additionally, for all BH masses, one is naturally limited by the velocity dispersion on the underlying stellar population in detecting the effect of a potential BH natal kick in the proper motion of a system.

Given all the above, and to explore the full range of plausible kicks imparted on newly born BHs and their effect on the properties of the BBH merger population, we consider two additional kick scenarios besides the default \texttt{No kick} scenario. In the \texttt{Normal kick} scenario, we use the natal kick velocity prescription from \citet{Disberg+25}, where, importantly, we do not scale the kick magnitude with the BH mass or the fallback fraction, as such scaling renders the kick effect negligible for most systems with $\Mone>39.7\,\Msun$. In this prescription, kick velocities are drawn from a log-normal distribution with $\mu\approx5.6$ and $\sigma=0.68$, with the kick direction drawn isotropically. We additionally implement a \texttt{Low kick} prescription, drawing from a log-normal distribution with $\mu=\log(40\: \mathrm{km/s})$ and $\sigma=0.68$ also without mass-rescaling. This lower mean is motivated by the radial velocity dispersion of \textit{Gaia} sources \citep{Yu+18, Anguiano+18}, such that natal kick signatures could not be distinguished from the observed velocity dispersion of the Galactic stellar population. 

\subsection{Population sampling}
For the populations at each metallicity, we sample $10^6$ ZAMS binaries from a \citet{Kroupa+01} initial mass function (IMF) between $7\Msun$ and $200\Msun$, a flat mass ratio between $0.05$ and $1$, and flat distribution in log separation between $5\Rsun$ and $10^5\Rsun$. These samples are reweighted to account for an IMF spanning $0.01{-}200\Msun$, a mass ratio between $0$ and $1$, and a binary fraction of 70\% \citep{Sana+12}. We convolve the metallicity-specific BBH mergers with the star formation and metallicity evolution of the IllustrisTNG-100 simulation \citep{Springel+18, Nelson+18, Pillepich+18, Naiman+18, Marinacci+18} and retain only BBH mergers with $\Mone>39.7\,\Msun$. For comparison between the simulated populations and inferred intrinsic population, we consider the three-dimensional \BGP model across $M_1$, $q$, and $\chieff$ from \citet{Abac+26a}. This model provides a non-parametric inference of the intrinsic BBH merger population from \texttt{GWTC-5.0}, and a marginal and joint posterior distribution over $\Mone$, mass ratio, and \chieff. We compare our predicted populations against the inferred distribution from the \BGP model at $z=0.2$, where the overall BBH merger population is best constrained \citep{Abac+25b}. For our simulated populations, we evaluate the average rate across the redshift range between $0.15\leq z \leq 0.25$. We discuss the redshift evolution of the BBH mergers from SMT in a separate work. In this work, the comparison is limited to the high-mass regime to specifically exclude the feature around ${\sim}35\Msun$. The boundary is set at $\Mone \geq 39.7\,\Msun$, following the bin edges uses based on in the \BGP model.

\section{Black hole accretion efficiency} \label{sec:results}


We compare the predicted \Mone distributions against the primary BH mass distribution inferred with a binned gaussian process (\BGP) on \texttt{GWTC-5.0} (gray region in Figure \ref{fig:m1_distributions}). We find that all three populations with \texttt{No kick} (left panel) predict merger rate densities above the observed rate. The Eddington-limited and GRRMHD populations, however, predict merger rate densities close to the inferred rate density, while the fully conservative population over-predicts the rate by more than an order of magnitude.

We find that the high-mass BBH merger populations is dominated by stable mass transfer, which accounts for ${\sim}95\%$ of events with $\Mone>39.7\Msun$. These systems originate from massive stars in low-metallicity stellar populations ($Z<\leq0.01\Zsun$). For more details on BBH formation through the stable mass transfer channel in \posydon, see \citet{Briel+26}. Contrary to \citet{Bavera+21} and \citet{vanSon+20}, we find that fully-conservative BH accretion does not significantly reduce the ability for systems to merge within a Hubble time (see Section \ref{sec:accretion} and Appendix \ref{app:super_eddington}). In our high-mass regime, fully-conservative accretion drastically increases the merger rate due to first-born BHs with $\Mone<39.7\,\Msun$ being able to gain sufficient mass to fall within the considered high-mass range. Because conservative mass transfer effectively shifts the primary BH mass distribution to higher masses, and low-mass BHs are the most abundant, the merger rate above ${\sim}40\Msun$ is drastically increased. Moreover, a smaller effect originates from the additionally retained mass, which result in more efficient gravitational wave emission, allowing for systems to merge that previously were too widely separated (see Appendix \ref{app:super_eddington}).
However, because the exact predicted merger rate in this regime is sensitive to the assumed initial conditions and high-redshift, low-metallicity star formation (Briel et al. in prep.), we limit the scope of this work to the general trend.

\subsection{\Mone distribution}

\begin{figure*}
    \centering
    \includegraphics[width=\linewidth]{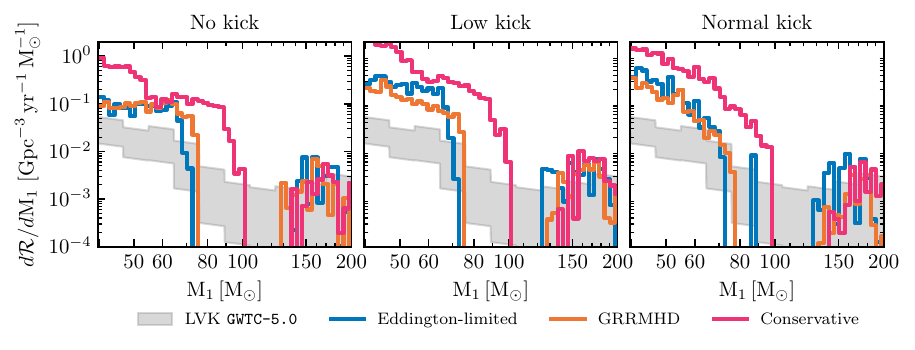}
    \caption{Primary BH mass distribution of BBH mergers with $0.15\leq z \leq 0.25$ and $\Mone>39.7\Msun$ for the Eddington-limited (blue), GRRMHD-informed (orange) and fully conservative (pink) BH accretion efficiency models. From left to right, the panel show populations with no kick, the low kick, and normal kick. The gray region is 95\% confidence interval of the \BGP model based on \texttt{GWTC-5.0} events with $\Mone \geq 39.7\Msun$ evaluated at $z=0.2$.}
    \label{fig:m1_distributions}
\end{figure*}


The left panel in Figure \ref{fig:m1_distributions} shows the predicted $\Mone$ distribution for the \texttt{No Kick} populations under the Eddington-limited (blue), GRRMHD-informed (orange), and fully conservative (pink) BH accretion prescriptions.
In this section, we only discuss the effects of the different mass accretion efficiencies. The effect of kicks and the remaining panels in Figure \ref{fig:m1_distributions} are deferred to Section \ref{sec:natal_kick}.

The primary BH mass distribution is divided into two regimes by the PISN mass gap. Above the upper PISN boundary at $\Mone \gtrsim 130\Msun$, all three prescriptions produce nearly identical distributions, despite some systems undergoing stable mass transfer. In Section \ref{sec:2D_distributions}, we show that this population can be separated into a subpopulation with low mass ratio ($q<0.4$) and a subpopulation with equal mass ratios ($q\sim 1$). Because the former subpopulation already undergoes a relatively efficient nuclear timescale mass transfer while on the main-sequence and the latter subpopulation undergoes chemically homogeneous evolution due to tidal spin up without interacting, we find that their outcomes are mostly insensitive to the BH accretion efficiency.

Below the PISN mass gap, the maximum primary BH mass increases from $\Mone=74\Msun$ for the Eddington-limited population to $\Mone = 77\Msun$ and $\Mone=101\Msun$ for the GRRMHD-informed and conservative populations, respectively. This trend is a direct effect of the increased BH accretion efficiency allowing the first-born BH to gain additional mass, and pollute the PISN mass gap. 

Although the inclusion of the hydrogen envelope in the remnant mass prescription allows for direct BH formation within the PISN mass gap, it does not continuously populate the gap between the lower and upper PISN boundary. The majority of BBH merger progenitors have lost their hydrogen envelope prior to collapse through mass transfer, as the high-mass ($\Mone \geq 39.7\Msun$) BBH merger population is dominated ($\geq 95\%$) by stable mass transfer. In Appendix~\ref{app:no_H}, we show that excluding hydrogen envelope fallback shifts the lower PISN boundary toward lower masses, confirming that its inclusion affects the location of this boundary.

The \texttt{No kick} populations in the left panel of Figure~\ref{fig:m1_distributions} show a nearly flat primary mass distribution between $39.7\Msun$ and ${\sim}70\Msun$, followed by a sharp decline toward their respective lower PISN mass gap boundaries. Because the GRRMHD-informed prescription exceeds the Eddington limit by only $10{-}30\%$, its predicted distribution closely resembles the Eddington-limited case across this range, with differences only near the PISN boundary. The predicted plateau and subsequent decline are broadly consistent with the tentative excess reported around $60{-}70\Msun$ in the \texttt{GWTC-4.0} population inference \citep{MaganaHernandez+24, Abac+25a}. However, this feature is largely absent in the more recent \texttt{GWTC-5.0} analysis \citep{Abac+26a}, suggesting that it may not yet be statistically robust or that the plateau is less pronounced than these models predict.

The fully conservative accretion prescription produces a steeper decline between $39.7\Msun$ and $90\Msun$ with a strong drop in merger rate around ${\sim}55\Msun$, and a sharp cut-off at ${\sim}100\Msun$. Neither feature is present in the \BGP model at these masses. Should future observations reveal structure above ${\sim}60\Msun$, it would provide direct evidence for the lower PISN boundary in the observed BBH population. None of these models, however, extend completely across the PISN mass gap itself, and they can, therefore, not account for BBH mergers across the full range of the primary masses.

\subsection{$q$ distribution}

\begin{figure*}
    \centering
    \includegraphics[width=\linewidth]{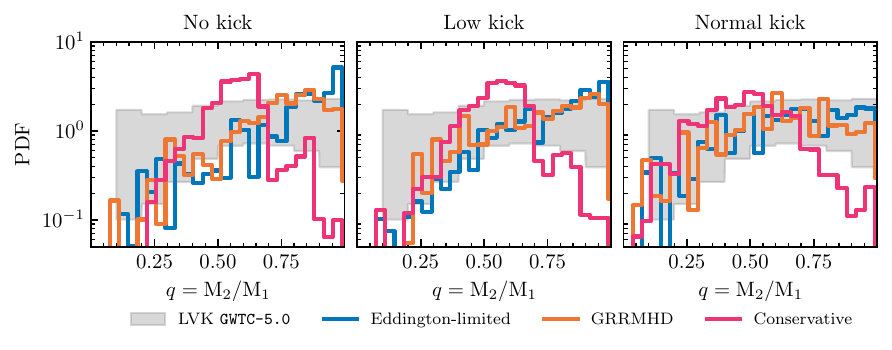}
    \caption{Mass ratio distributions for BBH mergers with $\Mone>39.7\Msun$ between $0.15 \leq z\leq 0.25$ shown for the \texttt{No kick} (left), \texttt{Low kick} (middle), and \texttt{Normal kick} (right) natal kick prescriptions, each evaluated with the Eddington-limited (blue), GRRMHD-informed (orange), and fully conservative (pink) BH accretion efficiencies. The gray shaded region denotes the \BGP inferred mass ratio distribution conditioned on $\Mone > 39.7\,\Msun$ and evaluated at $z=0.2$. The low and normal kicks are defined in Section~\ref{sec:methods}.}
    \label{fig:mass_ratio}
\end{figure*}

Although the first-born BH can increase more in mass as the accretion efficiency increases, the mass of the second-born BH remains unchanged. As such, the mass ratio distribution also shifts as the BH accretion efficiency increases. This effect can be seen in the left panel in Figure~\ref{fig:mass_ratio}, where the \texttt{No kick} Eddington-limited population with $\Mone>39.7\Msun$ shows a main peak from stable mass transfer at $q\approx 1$ with a tail towards lower mass ratios. The main peak shifts to $q\approx 0.85$ for the GRRMHD-informed accretion and to $q\approx 0.60$ for the fully conservative population. We find that the main mass ratio feature in the $M_1>39.7\Msun$ population is driven by stable mass transfer, and the location of this main mass ratio peak is shaped by the BH accretion efficiency in \posydon.

The inferred \BGP distribution for systems with $\Mone \geq 39.7\Msun$ has a broad mass ratio distribution \citep{Abac+26a}, which, in the \texttt{No kick} scenario, is more consistent with the Eddington limited and GRRMHD-informed populations than with the fully conservative case (left panel of Figure~\ref{fig:mass_ratio}). This is a consequence of the fully conservative scenario removing most equal mass ratio systems and shifting them to lower $q$ values, producing a clear peak at $q=0.6$ and dearth at equal masses in its distribution. Such a sharp peak and drop from the fully-conservative mass transfer is not supported by the data, thought a shallower peak such as those in the Eddington-limited and GRRMHD-informed population is supported.
Furthermore, in the high-mass regime, the observational sample remains small and contributions from multiple formation channels are likely \citep{Rodriguez+19, Kimball+20, Kimball+21, Borchers+25}. Distinguishing between accretion prescriptions and formation mechanisms on mass ratio alone will require a larger catalogue of high-mass BBH mergers, though stronger constraints can be reached in combination with other inferred parameters, such as spin. 

\subsection{\chieff distribution}

\begin{figure*}
    \centering
    \includegraphics[width=\linewidth]{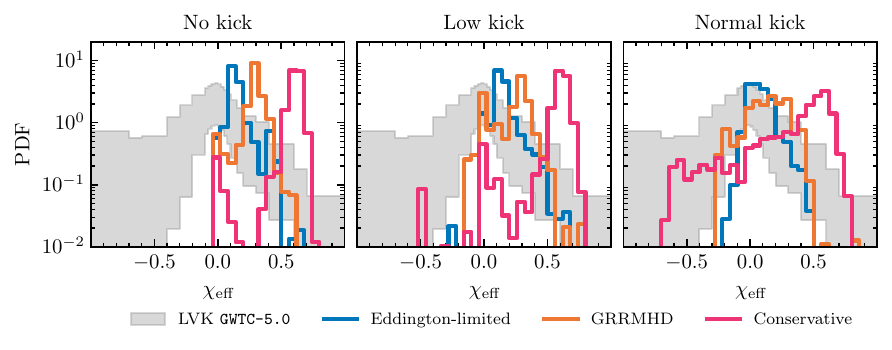}
    \caption{\chieff distributions for the different populations, as in Figure~\ref{fig:m1_distributions} and \ref{fig:mass_ratio}. The low and normal kicks are described in Section \ref{sec:methods}. The predicted BBH mergers are limited to $M_1>39.7\Msun$ and $0.15 \leq z \leq 0.25$. The \BGP model in gray is limited to the same mass range and evaluated at $z=0.2$.}
    \label{fig:chieff}
\end{figure*}

As the first-born BH accretes additional material, it also accretes angular momentum, spinning up and increasing its contribution to \chieff. The \texttt{No kick} scenario (left panel of Figure~\ref{fig:chieff}) most clearly shows the effect of accretion efficiency on the \chieff distribution. Since natal kicks are absent and the secondary spin is unaffected by the BH accretion efficiency, any differences in \chieff are, therefore, caused by the change in primary spin. Thus, as the BH accretion efficiency increases, \chieff increases, as is shown in the left panel in Figure \ref{fig:chieff}. The main peak in the fully conservative population is at $\chieff\approx0.6$, while for the GRRMHD-informed and Eddington-limited accretion prescription their peaks are at $\chieff\approx0.3$ and $\chieff\approx0.15$, respectively. The increased BH accretion during stable mass transfer clearly affects the final \chieff distribution. 

All three accretion efficiencies have an additional peak at \chieff=0, which originates from common envelope evolution. These systems have negligible accretion during the CE phase and retain a near zero spin. For the \texttt{No kick} fully conservative population, this produces a bimodal distribution in the \chieff distribution.

Observationally, the \BGP model suggests a preference for isotropy around $\chieff\sim0.0$, which requires support for negative \chieff. Due to the absence of a natal kick in the left panel in Figure \ref{fig:chieff}, no negative \chieff values are formed through isolated binary evolution, as expected. The main SMT \chieff peak from the Eddington-limited population aligns more closely to the potential peak in the \BGP distribution, while the peak of the fully-conservative model does not align well. This indicates that the angular momentum accreted by the first-born BH is limited, though this does not necessarily restrict the mass accretion.

\subsection{2-dimensional distributions} \label{sec:2D_distributions}

\begin{figure*}
    \centering
    \includegraphics[width=\linewidth]{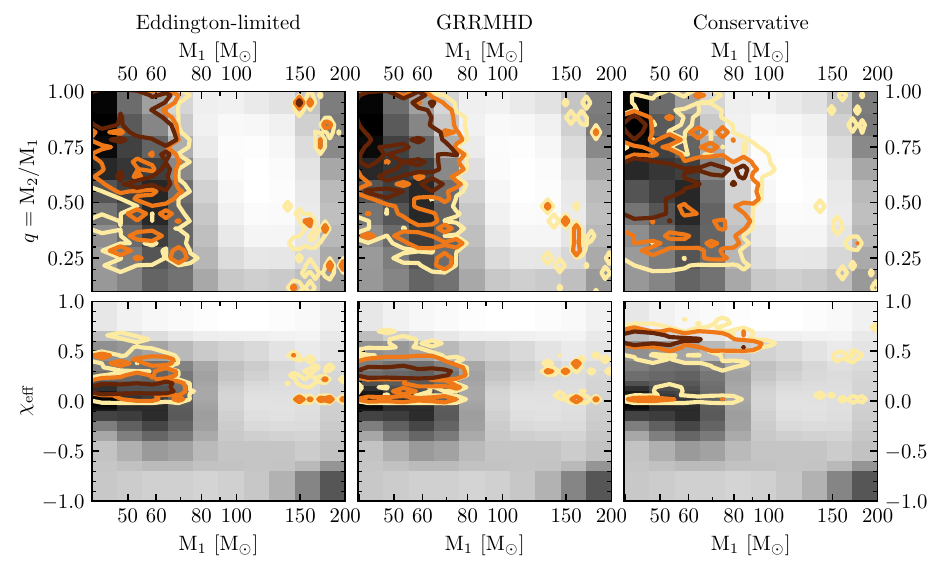}
    \caption{Two-dimensional density of properties of BBH mergers with $\Mone>39.7\Msun$ for Eddington-limited (left), GRRMHD-informed (middle), and conservative (right) BH accretion efficiencies in the interval $z = \left[0.15, 0.25\right]$ with no natal kick. The top row shows the mass ratio against primary mass with contours at 68\%, 95\%, and 99\% (dark red to light yellow). Additionally, the \BGP model from \texttt{GWTC-5.0} events is shown as the gray background 2-dimensional histogram. The bottom row shows $\chi_\mathrm{eff}$ against \Mone.}
    \label{fig:accretion_efficiency}
\end{figure*}

The previous sections have shown that varying the BH accretion efficiency leads to changes in \Mone, mass ratio, and \chieff distributions. Although these one-dimensional distributions provide crucial insight for comparison against observations, the BH accretion efficiency produces multi-dimensional correlations, where \Mone, mass ratio, and \chieff are being altered at the same time. Figure \ref{fig:accretion_efficiency} shows 2-dimensional distributions of mass ratio (top row) and \chieff (bottom row) against \Mone for the three \texttt{No kick} populations. The model predictions are shown as open contours and  the \BGP model as the gray 2-dimensional histogram in the background.

Across all three accretion populations, we make the distinction between four subpopulations in the model predictions, depicted as the open contours. These are most easily visible in the panel displaying \chieff against \Mone panel for the fully conservative population in the bottom right in Figure \ref{fig:accretion_efficiency}:
i) The population that has $\chieff\approx0$ with $\Mone \in [39.7, 90]\Msun$ and generally lower mass ratios ($q\simeq0.35$). These systems originate from common envelope evolution and are mostly unaffected by changes in the BH accretion efficiency. Comparing between the different panel in Figure \ref{fig:accretion_efficiency} shows that the conservative MT model allows for a slight increase in the maximum BH mass from this channel due to increased accretion onto the BH before the common envelope instability is triggered. However, this effect is limited on the final BH mass. 
ii) A similar population to i) with $\chieff\approx0$, but with above $\Mone > 150 \Msun$ and $q\in[0.7, 1]$. These systems have avoided an interaction due to evolving nearly chemically-homogeneous and are completely unaffected by the BH accretion efficiency. Because these systems are rare, the contours are sparsely populated.

The final two populations both originate from the SMT channel and are, thus, both affected by the BH accretion efficiency. Population iii) lies above the PISN mass gap with $\Mone>150\Msun$, the companion mass below the PISN mass gap with $q<0.4$, and relatively high $\chieff$ around $\simeq0.45$ in the bottom right panel of Figure \ref{fig:accretion_efficiency}. Because the mass and spin of the secondary is low, \chieff is dominated by the spin of \Mone.
The final population iv) is the main population with $\chieff>0.5$ in the bottom right panel of Figure~\ref{fig:accretion_efficiency} with $\Mone<110\Msun$. This population covers a large mass ratio range, but prefers $q\approx0.65$ in the fully-conservative population. This population is most affected by the BH accretion efficiency.  

A correlation between \chieff and $\Mone$ is present in the fully-conservative population: with increasing $\Mone$, the \chieff decreases slightly. This is a result of the additional accretion. To increase its spin to the same amount, an initially more massive BH needs to accrete more mass than a less massive BH. As a result, the lower $\Mone$ has slightly higher spins due to being able to achieve a higher spin with less mass and angular momentum accretion.

As described previously, an increasing BH accretion efficiency leads to an increasing \Mone mass, lower mass ratios, and higher \chieff, each of which can be seen in Figure \ref{fig:accretion_efficiency}.
In the same figure, the \BGP distribution, marked as the gray histogram, exhibits a drop in \Mone at $\sim 70\Msun$ with a preference for $q\simeq0.55$ and low \chieff. While the \BGP model may exhibit a peak in the mass ratio, the uncertainty is large enough for it to remain consistent with a flat mass-ratio distribution.
Comparing the \BGP distributions against the \texttt{No kick} populations, there is no isolated binary evolution model without a natal kick that produces the low \chieff, $q\sim0.55$ preference, and drop at $\Mone\sim70\Msun$ concurrently.
The fully conservative prescription provides a peaked mass ratio distribution at $q\sim0.65$ more closely aligning with the potential peak in mass ratio distribution at $q\simeq0.6$, and provides a high contribution of mergers above the lower PISN limit ($\Mone>70\Msun$). 
The Eddington-limited and GRRMHD-informed accretion efficiencies, on the other hand, align more closely with the potential decrease in $\Mone$ at $70\Msun$ and in inferred \chieff distributions, but produce a mass ratio distribution with a preference for $q\approx1$, though this effect is weakened by a natal kick, an effect which we will address in more detail in the following section.

\section{Natal kicks} \label{sec:natal_kick}

\subsection{\chieff distribution}
Although massive BHs are not expected to receive strong natal kick due to limited mass ejecta \citep{Janka+24, Burrows+25}, we, nevertheless, explore the effect of two different natal kick strengths on the merging BBH population. As described in Section \ref{sec:methods}, we have implemented a \texttt{Low kick} and a \texttt{Normal kick} prescription without any BH mass modulation. The inclusion of a natal kick allows the spins of the merging BHs to tilt into the orbital plane, introducing precession into the system, redistributing the dominant SMT peak towards lower \chieff values, and enabling negative \chieff values. The latter two effects are visible in the middle and right panels of Figure \ref{fig:chieff}, where the \texttt{Low kick} and \texttt{Normal kick}, respectively, have been applied to the populations. The prominent peaks present in the \texttt{No kick} population (left panel) are progressively smoothed out toward lower \chieff values and negative \chieff values as kick strength increases.

The fraction of systems with negative \chieff depends both on kick strength and BH accretion efficiency. For the \texttt{Low kick} model, the negative \chieff fractions are 3.5\%, 13.7\%, and 1.6\% for the Eddington-limited, GRRMHD-informed, and fully-conservative accretion efficiency, respectively. Under the \texttt{Normal kick} model, where the kick velocities are drawn from a log-normal distribution based on \citet{Disberg+25}, these fractions increase substantially, reaching 16.0\%, 18.8\%, and 12.6\% for the same accretion models. The degeneracy between kick strength and accretion efficiency nevertheless makes a direct interpretation of the observed negative \chieff fraction challenging, though a general increase in negative \chieff with stronger kicks is clear. The trend of a stronger natal kick leading to a higher fraction of negative \chieff is consistent with previous findings \citep[e.g.][]{Gerosa+18, Callister+21a, Banerjee+23, Baibhav+24}.

The choice of accretion prescriptions still determines the overall shape in the \texttt{Normal kick} scenario. The Eddington-limited prescription produces a distribution peaked near $\chieff \approx 0$ with a slight positive skew, while the GRRMHD-informed prescription yields a similarly shaped but moderately shifted and broader distribution peaking at $\chieff \sim 0.25$. The fully-conservative prescription, by contrast, retains a prominent peak at $\chieff \sim 0.6$. Although this feature is broadened and smoothed relative to the \texttt{No kick} population, it remains the dominant feature for the fully-conservative prescription.
The \BGP model shows a single peak near $\chieff\simeq0$ with a slight positive skew. Comparing this against the peaks of the predicted populations, the \texttt{Normal kick} Eddington-limited and GRRMHD-informed prescriptions both fall within the 95\% confidence interval of the \BGP model. The fully-conservative prescription, on the other hand, remains in tension with the inferred \chieff distribution, as its peak at $\chieff\sim0.6$ falls well outside the observed distribution.

\subsection{Mass ratio distribution}

A natal kick has a similar effect on the mass ratio, as it has on the \chieff distribution. As Figure \ref{fig:mass_ratio} shows, an increased natal kick strength leads to a smoothening of the peaks in the mass ratio distribution and allow more unequal mass ratio systems ($q<0.25$) to merge.

Under the \texttt{Low kick} scenario, the main peaks at $q\sim1$ for the the Eddington-limited and GRRMHD-informed prescriptions and at $q\sim0.64$ for the fully-conservative prescription are minimally broadened, leaving the main peaks largely intact.
The \texttt{Normal kick}, by contrast, produces for the Eddington-limited and GRRMHD-information prescriptions a nearly flat mass ratio distribution across the full mass ratio range, with only a gradual decline below $q\sim0.5$, while the fully-conservative prescription retains a preference for $q\sim0.5$, although broadened compared to the \texttt{No kick} case.

The strong peak in mass ratio of the fully-conservative population is in disagreement with the 95\% confidence interval of the mass ratio distribution from the \BGP model for all kick prescriptions. The inferred mass ratio distribution is mostly flat with a possible slight preference for more equal mass ratios. The Eddington-limited and GRRMHD-informed populations without \texttt{No kick} prefer too strongly equal mass ratios, but the mass ratio is flattened as the kick strength increases. The \texttt{Normal kick} scenario brings their mass ratio distribution into agreement with the inferred mass ratio distribution from the \BGP model.

\subsection{\Mone distribution}

For the primary BH mass distribution, increasing the natal kick strength progressively changes the shape of the distribution below the PISN mass gap, as shown in Figure \ref{fig:m1_distributions}. The plateau present in the \texttt{No kick} population gives way to a steeper power-law-like decline in the higher kick velocity models, though the transition from the power-law to the drop at the PISN mass limit becomes more gradual than in the \texttt{No kick} scenario. Although the \BGP model permits both a plateau or a slightly declining slope, the 95\% confidence interval shows a slight preference for a plateau, more closely aligned with a weaker natal kick scenario, though a higher statistical sample of BBH mergers is likely needed to distinguish between the power-law or plateau like drop-off. In the meantime, the mass ratio and \chieff distributions are likely to provide stronger constraints on the natal kick strength.

In the Eddington-limited population for the \texttt{Normal kick} models, systems are able to form and merge within the PISN mass gap, as the peak near $\sim90\Msun$ indicates. This is a consequence of non-interacting systems at low metallicities, where the complete hydrogen envelope falls back onto the newly formed BH, allowing BHs within the PISN mass gap to form during collapse. A strong natal kick is required for this channel to operate, as it necessitates an initially wide or non-interacting orbital configuration to preserve the hydrogen envelope. Under the \texttt{No kick} scenario, such systems are unable to merge within the Hubble time.  Interestingly, the upper mass limit of the PISN mass gap decreases with increasing natal kick strength, with systems of $\Mone \leq 150\Msun$ are able to merge within the Hubble time even under a weak kick. Depending on the natal kick strength non-interacting systems with a large hydrogen envelope are able to merger within the upper mass gap, allowing for the direct formation of mass gap BH, though in very limited numbers.

\subsection{2D dimensional distributions and summary}

\begin{figure*}
    \centering
    \includegraphics[width=\linewidth]{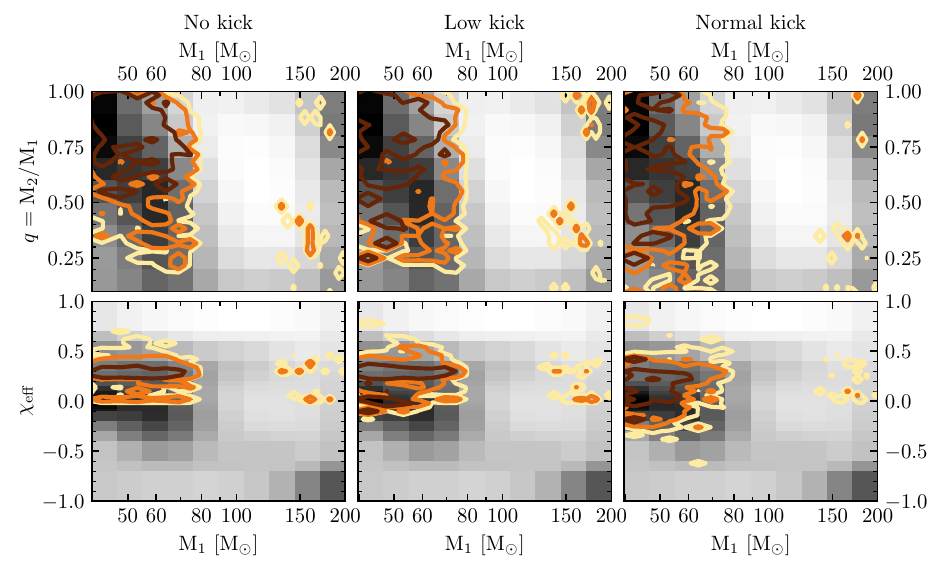}
    \caption{Predicted intrinsic population properties of BBH mergers with $0.15\leq z\leq 0.25$ and $\Mone\geq39.7\Msun$ from the GRRMHD BH accretion with no natal kick (left), a small natal kick (middle), and $\mu=270$km/s natal kick (right) populations. The top row shows the mass ratio against primary mass with contours at 68\%, 95\%, and 99\% (dark red to light yellow). Additionally, the \texttt{BGP} model from \texttt{GWTC-5.0} events is shown as the gray background 2-dimensional histogram \citep{Abac+25b}. The bottom row shows \chieff against \Mone.}
    \label{fig:natal_kicks}
\end{figure*}

For the GRRMHD-informed accretion model, we show the two-dimensional distributions for the three natal kick prescriptions in Figure \ref{fig:natal_kicks}.
Compared to the \texttt{No kick} populations in Figure \ref{fig:accretion_efficiency}, the main effects of increasing the natal kicks strength discussed in the previous section are visible: i) smoothening of the \chieff distribution and increased support for negative \chieff values, ii) greater support for more unequal mass ratios, and iii) a steepening of the \Mone slope below the PISN mass gap.

In the \chieff-\Mone plane (bottom row), the distinct subpopulation features present in the \texttt{No kick} population progressively blur together with increased kick strength, making it increasingly difficult to associate individual overdensities with specific formation channels. In the $q$-\Mone plane, no strong correlations between mass ratio and primary mass is present in any of the kick prescriptions. However, increasing the kick strength shifts the distribution towards lower mass ratios. It is clear that the combination of mass ratio and \chieff are most strongly affected by the natal kick strength. 

No strong correlations between the \Mone, mass ratio, or \chieff are present in the predicted populations. The \BGP model does not provide any strong evidence for or against correlations in the inferred population.

\section{Discussion} \label{sec:discussion}

Population inference suggests that the high-mass population likely contains a mixture of formation channels (see Section \ref{sec:introduction} and references there). This is further supported by the fact that isolated binary evolution with detailed binary models cannot, on its own, reproduce observations across all properties. The remainder of this section explores several implications of this picture.


\subsection{Additional features} \label{sec:additional_features}

So far we have primarily focused our analysis on the well-constrained parameters, \Mone, $q$, and \chieff, but our populations exhibit additional features in the individual spin components and their orientation. Even though these parameters are generally ill constrained with the current high-mass gravitational wave sample, our models exhibit several strong characteristics that might become observable as the sample increases in size. In particular, we highlight the primary spin magnitude $\chi_1$, and the effective precession spin, $\chi_\mathrm{p}$, which quantifies the degree of spin misalignment with the orbital angular momentum.

\begin{figure}
    \centering
    \includegraphics[width=\linewidth]{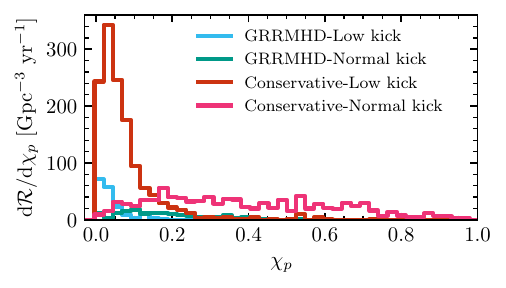}
    \caption{Intrinsic $\chi_p$ distribution for the super-Eddington accretion population with the \texttt{Low kick} and \texttt{Normal kick} prescription. Populations without the \texttt{No kick} scenario only have $\chi_\mathrm{p}=0$ and are therefore not plotted.}
    \label{fig:spin_precession}
\end{figure}

Figure \ref{fig:spin_precession} show the spin precession $\chi_\mathrm{p}$ in the kicked populations. Without a natal kick, $\chi_\mathrm{p}=0$ from isolated binary evolution and is, therefore, not shown in the Figure. A low natal kick already spreads out $\chi_\mathrm{p}=0$ towards higher values. For the GRRMHD-informed models, a peak at  $\chi_\mathrm{p}=0$ remains, while for the fully-conservative population, the $\chi_\mathrm{p}$ peaks away from zero for the \texttt{Low kick} scenario, in agreement with the models from \citet{Baibhav+24}. The \texttt{Normal kick} scenario further strengthens this effect, leading to a broader $\chi_\mathrm{p}$. Fully-conservative mass transfer leads to a nearly flat $\chi_p$ distribution between 0 and 0.8, after which is slowly decreased but to 1.0. The GRRMHD-informed population peaks at 0.075, but has events up to $\chi_p=0.5$. \citet{Ray+25a} reports a mostly flat $\chi_p$ distribution possibly peaking at $\chi_p\sim0.5$, though no clear distribution features are extracted from the observed events. Current observations are insufficient to make clear statements about the effective precession spin, but the models predicts strong features that will become detectable with future observations \citep{Vitale+25}.

\begin{figure}
    \centering
    \includegraphics[width=\linewidth]{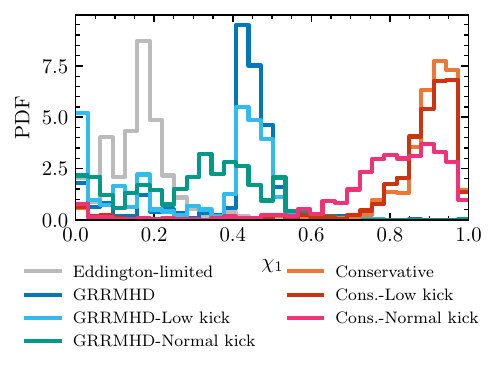}
    \caption{$\chi_1$ distributions for the Eddington-limited, GRRMHD, and fully-conservative BH accretion populations with different kick prescriptions for the GRRMHD and conservative populations. 
    We only show the \texttt{No kick} scenario for the Eddington-limited accretion efficiency, as the other kick scenarios follow a similar trend as the other BH accretion efficiencies, where lower $\chi_1$ are invoved in the merger. We show the distributions as PDFs due to the large rate differences between different populations.}
    \label{fig:chi1}
\end{figure}

Figure \ref{fig:chi1} shows the primary spin distribution. The fully-conservative models show a strong peak at $\chi_1\simeq0.9$ for the \texttt{No kick} and \texttt{Low kick} models, which becomes smoothed out towards lower $\chi_1$ in the \texttt{Normal kick} scenario. A similar behavior can be seen for the GRRMHD-informed models, but with their peak at $\chi_1 \simeq 0.45$ due to reduced mass and angular momentum accretion compared to the fully-conservative scenario.
Both GW190521 and GW231123 have measured high $\chi_\mathrm{p}$ ($0.68^{+0.25}_{-0.37}$ and $0.77^{+0.17}_{-0.19}$, respectively) with a low \chieff ($0.08^{+0.27}_{-0.36}$ and $0.31^{+0.24}_{-0.39}$) implying high primary BH spin magnitudes that are misaligned \citep{Abbott+20b, Abac+25c}. For our models to reproduce such high primary BH spin and high $\chi_\mathrm{p}$ events, the \texttt{Normal kick} with fully-conservative accretion model is required.
From population inference of the high-mass population, the distribution of primary BH spins is highly model dependent with \citet{Banagiri+25} finding a flat $\chi_1$ distribution, while \citet{Li+24a} reports a peaked distribution at 0.75. If the flat or high individual spin distribution can indeed be confirmed, a strong natal kick in isolated binary evolution is required to reconcile the low \chieff and high individual spin components in the super-Eddington accretion populations. 

In general, the spin of the secondary BH presents an even more uncertain picture with weak evidence for large $\chi_2$ values in specific events \citep{Abbott+20b, Abac+25c}. In all our models considered in this work, the secondary BH typically has a low spin of $\chi_2\sim 0.1$. \citet{Banagiri+25}, additionally, found some evidence for a transition in $\chi_2$ for the sub-population above $40\Msun$. This cannot be explained with super-Eddington accretion and a natal kick, and can point towards a different formation mechanism than isolated binary evolution. However, further investigation and high-mass events are needed to confirm this confidently, but would provide strong evidence for other formation mechanism than isolated binary evolution.

\subsection{On the strong natal kicks} \label{sec:high_natal_kick}

From isolated binary evolution, the tilt angles between the BH spins and the orbit is expected to peak at $\cos(\theta)=1$. However, several inference studies have pointed out that the tilt angles between the BH spins and the orbit of the full observed BBH merger population might point away from $\cos(\theta)=1$ \citep{Li+24a} or is flat \citep{Stegmann+25a, Galaudage+25}, though the data might currently be insufficient to draw strong conclusions on the spin tilt angle distribution \citet{Wolfe+26}.
Nevertheless, to reproduce spin tilts away from 1 ($\chi_p>0$) and get systems with negative \chieff in isolated binary evolution, a strong natal kick is required \citep{Baibhav+24}, which is supported by the models presented in this work.
This indicates that if the high-mass population indeed shows evidence for spin-orbit misalignment, isolated binary evolution requires BHs to receive a substantial natal kick at formation.

This requirement, however, is in tension with expectations from core-collapse supernova modeling. Massive BH progenitors are generally expected to collapse directly with only a small natal kick from asymmetric neutrino emission. This kick is typically a few km/s \citep{Rahman+22, Burrows+24}, rather than the dynamical kick associated with asymmetric mass ejections \citep[see][and references therein]{Janka+24}. While some 3D simulations find successful explosions and larger kicks for BH progenitors near the lower end of our mass range \citep[40\Msun;][]{Burrows+25}, most progenitors in our high-mass population arise from ZAMS masses where low natal kicks are expected \citep{Janka+24,Burrows+25}, though anisotropic infall from turbulent zones might induce a higher natal kick \citep{Antoni+23}. Despite the uncertainty in the 3D hydrodynamic modeling, it is unlikely that the high-mass BH progenitors will receive a strong natal kick during the core-collapse.

Our finding that a strong natal kick is required to reproduce the high-mass spin features (low \chieff, high $\chi_\mathrm{p}$, high $\chi_1$) from isolated binary evolution is, therefore, in tension with current core-collapse predictions for BHs in this mass range. Alternative mechanisms, such as asymmetric mass loss before collapse or during pair-instability pulsations \citep{Woosley+17}, or tossing of the spin axis during core-collapse \citep{Tauris+22}, could achieve similar spin features but these are not currently modeled in isolated binary evolution. Dynamical formation channels offer an alternative explanation for spin-orbit misalignment \citep[e.g.][]{Antonini+25}, and it is likely that contributions from multiple formation channels contribute to the high-mass BBH mergers \citep{Ray+26}. However, more GW merger events in the high-mass regime and multi-dimensional inference is required to confidently distinguish each contributing formation mechanism \citep{Zevin+21}.

\subsection{Comparison against other high-mass population synthesis results} \label{sec:accretion}

While there are many binary population synthesis work focusing on the complete mass range of BBH mergers \citep[see][and references therein]{Mandel+22a}, only a few have considered the population of BBH mergers above $\Mone\gtrsim40\Msun$. A majority of these altered the location of the PISN mass gap by altering the $^{12}C(\alpha, \gamma)^{16}O$ nuclear reaction rates to allow for more massive BH formation \citep{Farmer+19,Stevenson+19, Hendriks+23}. However, this still leaves a mass gap albeit it being at higher BH masses.
Super-Eddington accretion, on the other hand, pollutes the mass gap without altering the PISN physics. Previous studies focused on super-Eddington accretion in the context of massive BBH mergers find that super-Eddington accretion suppresses the formation of BBH mergers within the Hubble time \citep{vanSon+20, Bavera+21, Zevin+22}. The reduced angular momentum loss during the accretion leads to less orbital shrinkage compared to Eddington-limited accretion onto the BH. Thus, a progenitor systems must have tighter initial orbits to still merger within the Hubble time. Such systems will undergo Case A mass transfer, which is highly uncertain in rapid population synthesis codes, often not leading to a BBH mergers \citep[e.g.][]{Romero-Shaw+20, Broekgaarden+22}. As a result, these studies conclude that fully conservative accretion keeps BBH progenitors too wide to merge, substantially reducing the total merger rate and making it difficult to populate the upper mass gap \citep{vanSon+20}. In contrast, using detailed binary models, \citet{Briel+23} find that BBH mergers do form even under fully conservative accretion.

In this work, we confirm that super-Eddington accretion does not lead to a reduction in the BBH merger rate, but even an increase in BBH mergers (see also Appendix~\ref{app:super_eddington}). This differences arises because, in detailed binary models, Case A mass transfer is the dominant formation pathway for BBH mergers through the SMT channel \citep{Briel+26}, rather than being a source of failed mergers as in rapid population synthesis. The majority of massive BBH mergers in \posydon originate from low-metallicity, short orbital-period Case A systems, where the structure of the donor determines the moment the system detaches during the STAR+BH mass transfer \citep{Briel+26}. Because these Case A progenitors are already tight  at ZAMS, they do not require the additional angular momentum loss from Eddington-limited accretion to shrink the orbit sufficiently to merger within the Hubble time. This is different from the initially wider Case B systems that make up the majority of BBH merger in rapid population synthesis.
As a consequence, the accretion efficiency has a limited impact on the post-mass transfer orbital configuration \citep[see Appendix~\ref{app:super_eddington} and][]{Klencki+26}, though it still affects the merger time through its effect on gravitational-wave emission. As a result, super-Eddington accretion onto the BH can still produce BBH mergers with primary masses inside the PISN mass gap, without the significant rate suppression found in earlier work.

\subsection{The rate density and star formation history} \label{sec:rate_density}

While the previous analysis has focused on the shape of the distributions, here we consider the intrinsic rate density of BBH mergers with $M_1\geq 39.7\,\,\Msun$. In this mass range, the \BGP model infers a rate of $0.57^{+0.8}_{-0.33}$ events per Gpc$^{3}$ per year at $z=0.2$. As shown in Table \ref{tab:rates}, all models in this work overpredict the merger rate density. Even the lowest merger rate at $2.7$ event per Gpc$^{3}$ per year in \texttt{No kick} scenario with Eddington-limited accretion exceeds the upper boundary of the \BGP model, and this model cannot concurrently explain the other properties of the merging population.

This overprediction from isolated binary evolution bove $\Mone>39.7\Msun$ can, however, be mitigated by uncertainty in the star formation history (Briel et al., in prep). The high-mass BBH merger population originates predominantly from low-metallicity environments ($Z<0.01\Zsun$), and is characterized by long delay times \citep[for more details, see][]{Briel+26}. This makes the local intrinsic rate density sensitive to low-metallicity star formation at high redshift, which is poorly constrained by observations.  Briel et al.\ (in prep) show that the high-mass tail of the primary BH mass distribution can be reduced accordingly, allowing for additional contribution from other formation channels in the high-mass primary BH mass regime.





\section{Conclusion} \label{sec:conclusion}

In this work, we have investigated the formation of BBH mergers with primary BH masses above $39.7\,\Msun$ through isolated binary evolution, exploring whether super-Eddington accretion during SMT can explain the observed population of massive BBH mergers near and within the upper PISN mass gap. Using the \posydon binary population synthesis code, we examined the impact of varying BH accretion efficiencies and natal kick prescriptions to determine if isolated evolution can reproduce the multi-dimensional properties of the current gravitational wave observations of the high-mass BBH merger population.
 
We find that when BHs are able to accrete at super-Eddington rates, isolated binary evolution can produce BBH mergers with primary BH masses that extend into the PISN mass gap, and that these systems are able to merge within the Hubble time. In contrast to previous studies \citep[e.g.][; see Section \ref{sec:accretion} for details]{vanSon+20, Bavera+21}, we find that BBH merger rates are not suppressed under super-Eddington accretion. This differences arises because detailed binary models shift the dominant mass-transfer from post-main-sequence to main-sequence mass transfer, requiring less orbital shrinkage to merge within the Hubble time \citep[see][]{Briel+26}.
In the case of fully-conservative BH accretion, this shifts the maximum BH mass below the PISN mass gap from ${\sim}60\Msun$ to ${\sim}100\Msun$, while still allowing systems to merge within the Hubble time. While these models are able to reproduce some of the characteristics of the observed high-mass BBH merger population, they cannot match all of them concurrently.

In general, more efficient BH accretion leads to BBH mergers with higher primary BH masses, more unequal mass ratios, and higher \chieff values. However, super-Eddington accretion does not populate the complete upper mass gap.
The GRRMHD-informed and Eddington-limited prescriptions best reproduce the inferred \chieff, mass ratio, and \Mone distributions from the \BGP model from \texttt{GWTC-5.0} above $39.7\Msun$. However, in the absence of natal kicks, both produce strong narrow features in \chieff and mass ratio that are broader in the inferred population. The fully-conservative population, while able to populate the PISN mass gap, has a strong peak at $\chieff\sim0.6$ and $q\sim0.6$. Populations without a natal kick universally produce strong features in \chieff and mass ratio, demonstrating that natal kicks are required to reproduce the full morphology of the inferred observational population, if the mergers originate from isolated binary evolution.

We explored two natal kick strengths, without BH mass modulation, and find that increasing kick strength changes the BBH merger properties in distinct ways:
i) The \Mone distribution below the PISN mass gap steepens from a plateau into a power-law decline, with a more gradual transition into the lower PISN limit compared to the \texttt{No kick} case.
ii) The upper PISN limit decreases in the presence of a natal kick, with more systems of $\Mone \lesssim 150\Msun$ being able to merge within the Hubble time.

Additional, we find that a strong natal kick combined with the fallback of the hydrogen envelope during core-collapse at low metallicities, allows for the formation of BBH mergers within the PISN mass gap without super-Eddington accretion rates. The natal kick further flattens the mass ratio distribution and increases support for unequal mass ratio systems. Furthermore, it broadens the \chieff distribution toward lower and negative values with the fraction of negative \chieff systems scaling with both kick strength and accretion efficiency, which is most efficient (20.7\%) with a strong kick (the \texttt{Normal kick} model) and GRRMHD-informed BH accretion.
Whether the most massive BHs can indeed receive strong natal kicks remains an open question beyond the scope of this work; we refer the reader to Section \ref{sec:discussion} for further discussion.

Comparing the predicted populations to the \BGP model, the GRRMHD-informed and Eddington-limited prescriptions best reproduce the inferred \chieff, mass ratio, and \Mone distributions, provided a strong natal kick is applied. Both models produce a decrease in merger rate near $\sim 70\,\Msun$, consistent with a potential feature in the \BGP \Mone distribution. A strong natal kick also facilitates reconciliation with the low effective spin and high precession signatures inferred from GW observations, if these are confirmed. The fully-conservative prescription requires a strong natal kick to achieve compatibility in mass ratio, but even then retains a prominent peak at $\chieff\sim 0.6$ that remains in tension with the \BGP distribution.

Despite being able to match individual features in the mass distribution, isolated binary evolution is unable to produce a continuous primary BH mass distribution that reaches through the PISN mass gap for any of the explored accretion efficiencies, which points to additional formation channels or the absence of PPI and PISN. Distinguishing between the plateau and power-law morphologies in the \Mone distribution will likely require a larger GW event sample. In the meantime, the joint \chieff and mass ratio distributions provide a possible avenue to distinguish between accretion prescriptions and kick strengths. However, the current observed population of high-mass BBH mergers is insufficient to isolate the isolated binary evolution contribution to the high-mass regime. Future observations are likely to enable such inference and place stronger constraints on both BH accretion physics and natal kick prescriptions in the massive star regime.

\begin{acknowledgements}

The \posydon project is supported primarily by two sources: the Swiss National Science Foundation (PI Fragos, project number CRSII5\_213497) and the Gordon and Betty Moore Foundation (PI Kalogera, grant award GBMF8477). MZ gratefully acknowledges funding from the Brinson Foundation in support of astrophysics research at the Adler Planetarium. J.J.A.\ acknowledges support for Program number (JWST-AR-04369.001-A) provided through a grant from the STScI under NASA contract NAS5-03127. A.R. was supported by the National Science Foundation~(NSF) award PHY-2512923. The computations were performed at Northwestern University on the Trident computer cluster (funded by the GBMF8477 award) and at the University of Geneva on the Yggdrasil computer cluster. This research was supported in part through the computational resources and staff contributions provided for the Quest high performance computing facility at Northwestern University which is jointly supported by the Office of the Provost, the Office for Research, and Northwestern University Information Technology.

\end{acknowledgements}

\bibliographystyle{aasjournalv7}
\bibliography{high-mass_BBH_mergers}

\appendix

\section{Not conserving the hydrogen envelope} \label{app:no_H}

\begin{figure*}
    \centering
    \includegraphics[width=\linewidth]{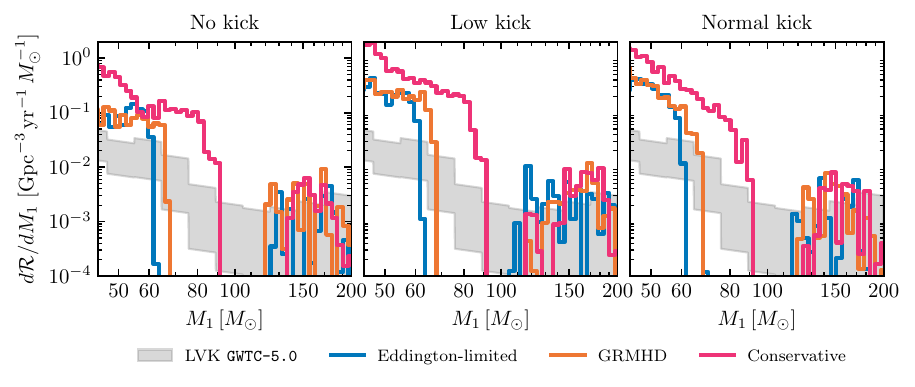}
    \caption{Primary BH mass distributions of BBH mergers with $0.15\leq z \leq 0.25$ and $\Mone>39.7\,\Msun$ for the Eddington-limited (blue), GRRMHD-informed (orange), and fully conservative (pink) BH accretion efficiency models. From left to right the panels show populations with \texttt{No kick}, \texttt{Low kick}, and \texttt{Normal kick} prescriptions. The gray regions indicated the 95\% confidence interval of the \BGP model based on \texttt{GWTC-5.0} events with $\Mone>39.7\,\Msun$. These populations do not include the hydrogen envelope during direct-collapse.}
    \label{fig:m1_distributions_no_H}
\end{figure*}

Conserving the hydrogen envelope during direct-collapse allows for additional material to fall back onto the BH, increasing the final BH mass. Whether this material is retained in nature through the final phases of stellar evolution and available during collapse remains uncertain \citep[see for example][]{Chan+18}. Nevertheless, we included the hydrogen-conserved scenario in our main results, because it allows us to more easily produce more massive BHs which potentially populate the PISN mass gap.

Excluding the hydrogen envelope from the remnant mass calculation leads no perceivable difference in \chieff distributions and produces only minimal changes to the mass ratio distribution. The main effect is on the $\Mone$ distribution, shown in Figure~\ref{fig:m1_distributions_no_H}, where the maximum BH mass below the PISN mass gap is decreased compared to the hydrogen included case. It reaches $\sim65\Msun$ for Eddington-limited, ${\sim}70\Msun$ for GRRMHD-informed, and ${\sim}95\Msun$ for fully-conservative accretion prescription.

In the hydrogen-conserved scenario, the Eddington-limited and GRRMHD-informed accretion prescriptions produced similar maximum BH masses below the PISN mass gap compared to the hydrogen-exclusion scenario. The systems that set the maximum BH mass originate from progenitors just below the PPI threshold that retain a large hydrogen envelope, and because of the additional fallback material produces a massive BH. In the hydrogen-exclusion scenario, the massive BHs come from the same regime, but do not have the additional mass that falls back onto the BH, thus producing lower mass limits of the lower PISN boundary. The turn-off features in \Mone for the Eddington-limited and GRRMHD-informed prescriptions coincide with the possible features in the \BGP model, though at slightly lower masses.

The upper boundary of the PISN mass gap is also affected by the inclusion of the hydrogen envelope. In the no-kick population, the limit shifts to ${\sim}110\Msun$ compared to ${\sim}140\Msun$ in the hydrogen-conserved scenario for all BH accretion efficiencies. Furthermore, the effect from natal kicks on the shift of this upper boundary is decreased in the hydrogen-exclusion scenario with the lower mass systems already merging in the \texttt{No kick} population.
The plateau structure visible below the PISN limit is, however, still affected by the natal kick strength, which shifts is into a steeper power-law distribution.

\section{The effect of BH accretion efficiency on the \texttt{CO-HMS\_RLO} grid} \label{app:super_eddington}

\begin{figure*}
        \centering
        \includegraphics[width=\linewidth]{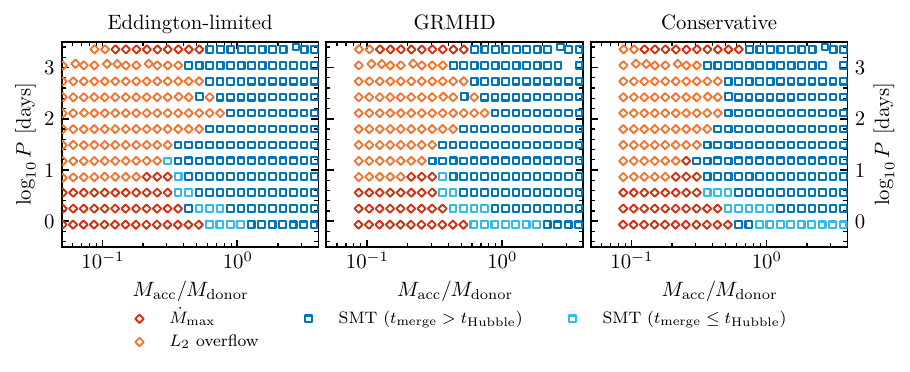}
    \caption{Slices of the \texttt{CO-HMS\_RLO} \mesa grids at $Z=10^{-2}\, \Zsun$ with $M_\mathrm{donor}\approx 33\,\Msun$ with Eddington-limited (left), GRRMHD-informed (middle), and fully conservative (right) BH accretion. The diamond markers in red and orange indicate models reaching the unstable mass transfer conditions of the maximum mass transfer rate of 0.1\Msun/yr and $L_2$ outflow, respectively. Stable mass transfer models are indicates with square markers, where systems with a final orbital configuration that merge within the Hubble time are marked in light blue while non-merging systems are marked in dark blue. The period and $M_\mathrm{acc}/\Mdonor$ are at the start of Roche lobe, since the \texttt{CO-HMS\_RLO} grid starts there. This causes the models in the top-right to not follow the initial grid paramters in the top-right.} 
    \label{fig:CO-HMS_RLO_grids}
\end{figure*}

Previous works \citep{vanSon+20, Bavera+21, Zevin+22} have suggested that super-Eddington accretion cannot significantly populate the PISN mass gap, because the mass and angular momentum lost through inefficient BH accretion are required to shrink the orbit efficiently for the BBH system to merge within the Hubble time. 

Two competing effects determine whether a system can merge within the Hubble time: i) the orbital period after the formation of the secondary BH, and ii) the masses of the BHs, which determine the efficiency of gravitational wave emission. The relative importance of these effects changes as the BH accretion efficiency increases and their interplay determines whether the BBH merger rate is suppressed or enhanced. 

These effects play an important role in different regions of the \texttt{CO-HMS\_RLO} grid. Figure \ref{fig:CO-HMS_RLO_grids} shows the three grid slices of $\mathrm{M}_\mathrm{donor}=33\Msun$ at $0.1\Zsun$ for the Eddington-limited (left), GRRMHD-informed (middle), and fully conservative (right) BH accretion efficiencies. Increasing the accretion efficiency from Eddington-limited to the GRRMHD-infrmed prescription, reduced the mass lost from the system and therefore reduces the orbital shrinkage. This makes it more difficult for systems to merge within the Hubble time, qualitatively consistent with \citet{vanSon+20} and \citet{Bavera+21}, though the effect is less pronounced in our models. It is most apparent in wide period region leading to a BBH merger through SMT (light blue) at ${\sim}16$ days, where the widest light blue marker in the Eddington-limited grids is replaced by a non-merging dark blue marker in the GRRMHD-informed grids. In the fully conservative grid slice, excessive orbital widening arises due to mass ratio reversal, preventing systems from merging entirely.

However, systems that are initially on tight orbits are less sensitive to changes in orbital separation. Because more mass is retained within the system under higher accretion efficiencies, these compact systems can merge more efficiently within the Hubble time through enhanced gravitational wave emission. In the GRRMHD-informed prescription, however, the BH accretion efficiency is not high enough to substantially expand this parameter space. This is visible in Figure \ref{fig:CO-HMS_RLO_grids} at the tightest SMT systems, where the SMT-to-BBH-merge region, the light blue region, is only slightly expanding. Moving to the fully-conservative accretion, this initially tight orbit leading to merger expands substantially. While the widest systems no longer merge, the additional parameter space to merger from SMT at the tightest orbits produces a net increase in the overall BBH merger rate.

As a result of this interplay, we do not find a suppression of the BBH merger rate density with increasing BH accretion efficiency. On the contrary, the number of massive BBH mergers increases. Although this can partly be attributed to lower-mass systems gaining sufficient mass to enter the $\Mone > 39.7\, \Msun$ regime, the overall BBH merger rate is not suppressed by super-Eddington accretion.

As in the high-mass population from \texttt{BPASS} \citep{Briel+23}, fully conservative BH accretion still permits BBH systems to merge within the Hubble time in \posydon. Because the \posydon SMT BBH mergers primarily undergo main sequence mass transfer (Case A) and require initially tight orbits to merge within the Hubble time \citep{Briel+26}, the effect of orbital widening is limited. As shown in Figure~\ref{fig:CO-HMS_RLO_grids}, the longest-period systems around $10$ days in the SMT-to-BBH-merge region (light blue squares) are removed as BH accretion efficiency increased from left to right. In BSE-based codes, BBH merger progenitors predominantly undergo Case B mass transfer and occupy these and wider orbits, likely explaining why BH accretion efficiency more strongly affects merger rates in those population synthesis models. By modeling the binaries in detail, however, we are find that Case A is the dominant interactions leading to BBH mergers in the SMT channel, and leads to a more complex dependence of the BBH merger rate on the BH accretion efficiency.

\subsection{Table of the predicted BBH merger rate density}

\begin{table}[]
    \centering
    \begin{tabular}{c|c|c|c}
         & \multicolumn{3}{c}{Rate Density [Gpc$^{-3}$ yr$^{-1}$]} \\
         \hline
         & \texttt{No kick} & \texttt{Low kick} & \texttt{Normal kick} \\
         \hline
        \textbf{Eddington-limited} &  2.7  & 6.9 & 6.2 \\
        \textbf{GRRMHD-informed} &  3.1 & 4.3  & 3.8 \\
        \textbf{Fully conservative} & 12.0 & 31.8 & 23.2 \\
        \hline
    \end{tabular}
    \caption{BBH merger rate density between $0.15\leq z \leq 0.25$ for systems with $\Mone>39.7\Msun$.}
    \label{tab:rates}
\end{table}

\end{document}